\documentclass[aps,prl,onecolumn,groupedaddress,notitlepage,10pt]{revtex4-1}

\usepackage{upgreek}
\usepackage{graphicx}
\usepackage[colorlinks=true,urlcolor=black,linkcolor=black,citecolor=black]{hyperref}
\usepackage[all]{hypcap} 
\usepackage[labelfont=bf]{caption}
\usepackage{color}
\usepackage{pdfpages}
\usepackage{subfiles}

\begin{document}

%\preprint{}

\title{Tabletop Nanometer Extreme Ultraviolet Imaging in an Extended Reflection Mode using Coherent Fresnel Ptychography}


\author{Matthew~D.~Seaberg*}
%\email[]{mattseaberg@gmail.com}
\author{Bosheng~Zhang}
\author{Dennis~F.~Gardner}
\author{Elisabeth~R.~Shanblatt}
\author{Margaret~M.~Murnane}
\author{Henry~C.~Kapteyn}
\author{Daniel~E.~Adams}
%\thanks{}
\affiliation{JILA, University of Colorado at Boulder, Boulder, CO 80309, USA\\
*\href{mailto:mattseaberg@gmail.com}{\color{blue} mattseaberg@gmail.com}}

%\collaboration{Matthew D. Seaberg}

%\date{\today}

\begin{abstract}
We demonstrate high resolution extreme ultraviolet (EUV) coherent diffractive imaging in the most general reflection geometry by combining ptychography with tilted plane correction. This method makes it possible to image extended surfaces at any angle of incidence. Refocused light from a tabletop coherent high harmonic light source at 29 nm illuminates a nanopatterned surface at $45^\circ$ angle of incidence. The reconstructed image contains quantitative amplitude and phase (in this case pattern height) information, comparing favorably with both scanning electron microscope and atomic force microscopy images. In the future, this approach will enable imaging of complex surfaces and nanostructures with sub-10 nm-spatial resolution and fs-temporal resolution, which will impact a broad range of nanoscience and nanotechnology including for direct application in actinic inspection in support of EUV lithography.
\end{abstract}

%\pacs{}
%\keywords{}

\maketitle


Dramatic advances in coherent diffractive imaging (CDI) using light in the extreme ultraviolet (EUV) and X ray regions of the spectrum over the past 15 years have resulted in near diffraction-limited imaging capabilities using both large and small scale light sources \cite{Miao1999,Seaberg2011}. In CDI, also called ``lensless imaging,'' coherent light illuminates a sample, and the scattered light is directly captured by a detector without any intervening imaging optic. Phase retrieval algorithms are then applied to the data set to recover an image. CDI has already been used to study a variety of biological and materials systems \cite{Jiang2010,Nelson2010,Clark2013,Turner2011,Xu2013}. However, the potential for harnessing the power of CDI for imaging complex nano-structured surfaces, which requires the use of a reflection geometry for imaging, has been much less studied. Surfaces are critical in nanoscience and nanotechnology, for example in catalysis, energy harvesting systems or nanoelectronics. A few successful demonstrations have applied CDI to reflection-mode imaging. However, work to date has either been limited to highly reflective EUV lithography masks in a normal incidence geometry \cite{Harada2013}, restricted to low numerical aperture through the use of a transmissive mask \cite{Roy2011}, or restricted to isolated objects \cite{Zurch2013,Sun2012}.

Here we demonstrate the most general reflection-mode coherent diffractive imaging to date using any light source, by combining the extended ptychographical iterative engine (ePIE) \cite{Maiden2009} with curved wavefront illumination \cite{Vine2009}. This allows extended (non-isolated) objects to be imaged at any angle, which will enable tomographic imaging of surfaces. This work also represents the first non-isolated-object, high fidelity, tabletop coherent reflection imaging, which expands the scope of applications for CDI significantly. This work demonstrates a powerful new capability that can impact a very broad range of science and technology. First, our approach removes restrictions on the numerical aperture, sample, or angle, so that general extended objects can be imaged in reflection mode at any angle of incidence. Second, illumination of the sample with a strongly curved wavefront removes the need for a zero-order beam-stop by reducing the dynamic range of the diffraction patterns. The curved illumination also allows the size of the beam to vary according to the sample size, alleviating the need for a large number of scan positions. This also results in fewer necessary scan positions when imaging a large field of view. Third, reflection ptychography produces surface images containing quantitative amplitude and phase information about the sample that are in excellent agreement with atomic force microscopy (AFM) and scanning electron microscopy (SEM) images, and also removes all negative effects of non-uniform illumination of the sample or imperfect knowledge of the sample position as it is scanned. The result is a general and extensible imaging technique that can provide a comprehensive and definitive characterization of how light at any wavelength scatters from an object, with no fundamental limitation on resolution. This complete amplitude and phase characterization thus is fully capable of pushing full field optical imaging to its fundamental limit. Finally, because we use a tabletop high harmonic generation (HHG) 30 nm source \cite{Popmintchev2012}, in the future it will be possible to image energy, charge and spin transport with nm spatial and fs temporal resolution on nanostructured surfaces or buried interfaces, which is a grand challenge in nanoscience and nanotechnology \cite{Nardi2011,Mathias2012}.

\begin{figure}[thb]
\includegraphics[width=6.in]{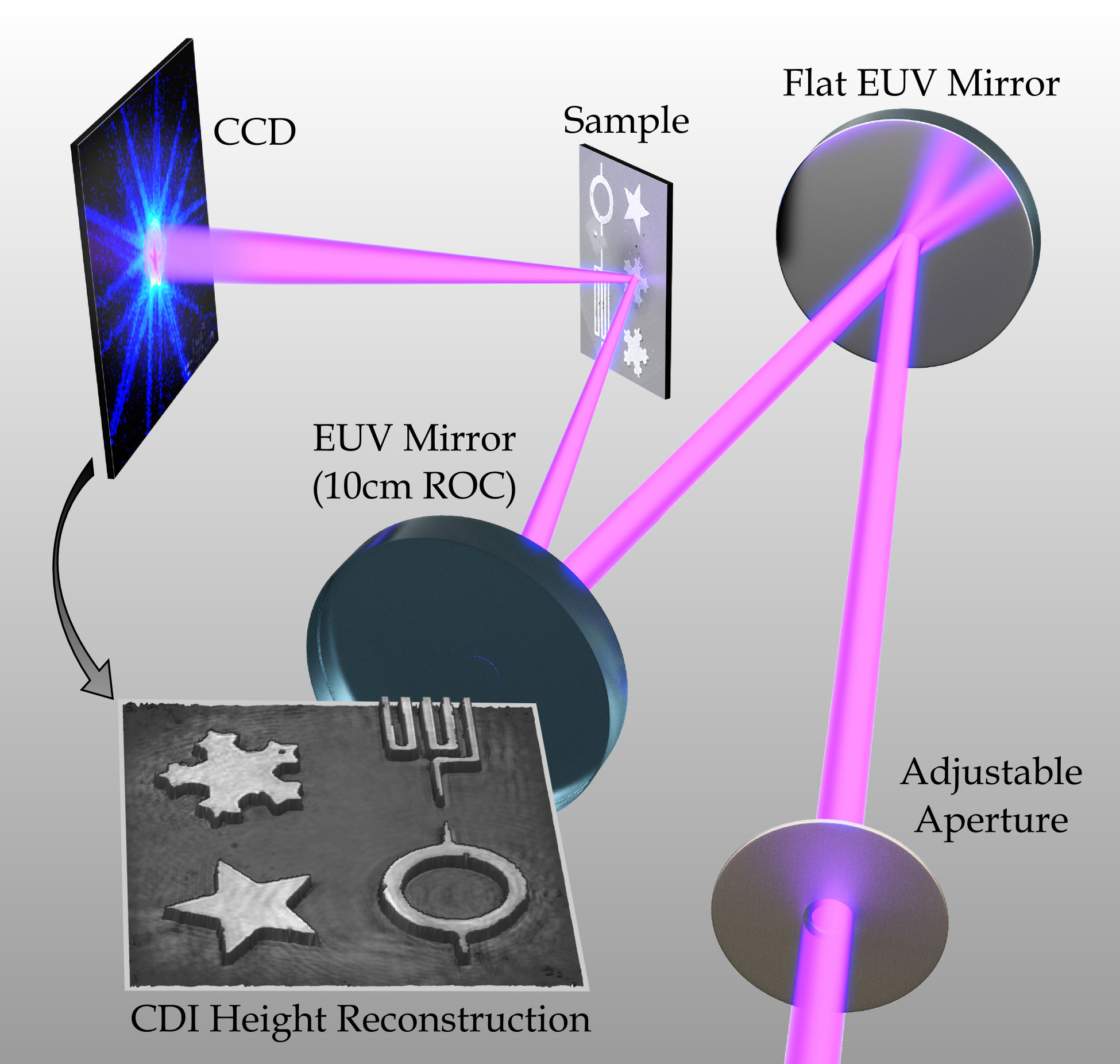}
\caption {{\bfseries Experimental setup for reflection mode Fresnel ptychography.} The EUV beam propagates through an adjustable $\approx$1 mm aperture, and a single harmonic is selected using a pair of multilayer mirrors centered at 29.5 nm, and focused onto the sample. The scattered light is collected on a CCD detector placed directly after the sample. The inset shows a height profile reconstructed through ptychography.}
\label{schematic}
\end{figure}

The experimental geometry for reflection mode Fresnel ptychography is shown in Fig.~\ref{schematic}. A Ti:sapphire laser beam with wavelength $\approx$785 nm (1.5 mJ pulse energy, 22 fs pulse duration, 5 kHz repetition rate) is coupled into a 5 cm-long, 200 ${\rm {\upmu}m}$ inner diameter, hollow waveguide filled with 60 torr of argon. Bright harmonics of the fundamental laser are produced near a center wavelength of 29 nm (27th harmonic) since the high harmonic generation process is well phase-matched \cite{Bartels2002}, ensuring strong coherent signal growth and high spatial coherence. The residual fundamental laser light, which is collinear with the high harmonic beam, is filtered out using a combination of two silicon mirrors (placed near Brewster’s angle for 785 nm light) and two 200 nm-thick aluminum filters. The EUV beam is then sent through an adjustable $\approx$1 mm aperture, placed $\approx$1 m upstream of the sample, to remove any stray light outside the beam radius. A pair of Mg/SiC multilayer mirrors then select the 27th harmonic of the Ti:sapphire laser at 29.5 nm. The first mirror is flat, while the second mirror has a radius of curvature of 10 cm. This mirror pair focuses the HHG beam onto the sample at an angle of incidence of $45^\circ$. The actual focus is 300 ${\rm {\upmu}m}$ behind the sample, so that the HHG beam wavefront has significant curvature. The angle of incidence on the curved mirror is approximately $2^\circ$, which introduces small amounts of astigmatism and coma onto the HHG beam.
 
\begin{figure}[htb]
\includegraphics[width=6.in]{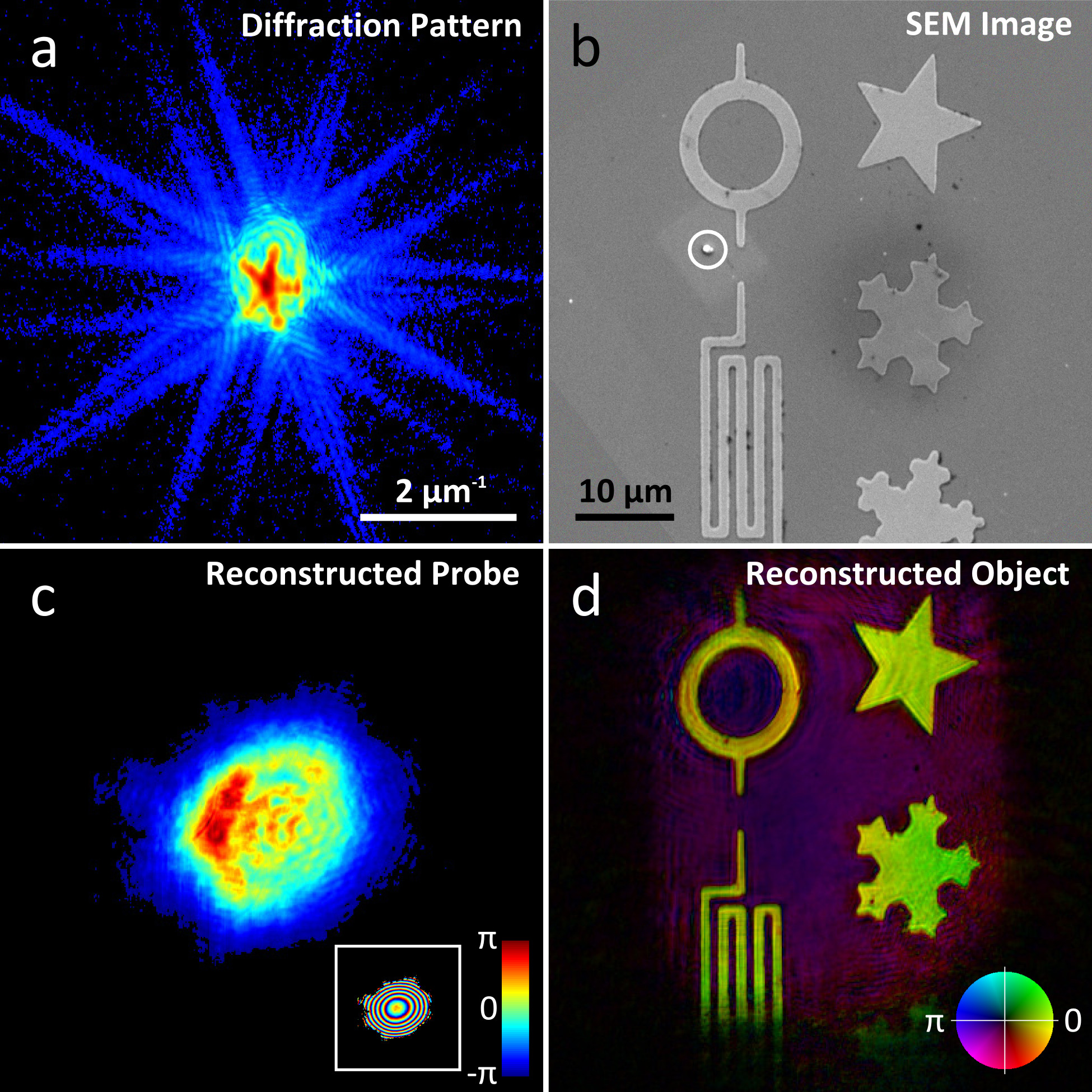}
\caption{{\bfseries Diffraction data and ptychographic reconstruction.} (a) Representative diffraction pattern, scaled to the $\frac{1}{4}$ power, taken from the 90-scan dataset. (b) SEM image of the Ti patterned Si sample. Note that the large defect circled in the SEM image resulted from contamination after the ptychography measurement. (c) Reconstructed amplitude (thresholded at 5$\%$) of the HHG beam. The inset shows the reconstructed phase (displayed modulo-2$\uppi$). (d) Ptychographic reconstruction of the object shown in (b). The reconstruction is plotted as the complex amplitude, where brightness represents reflected amplitude and hue represents the phase of the reconstruction. Note that the majority of defects seen in the SEM image of the Ti nanostructures are reproduced in the ptychographic reconstruction. The scale bar in (b) is shared among (b)-(d).}
\label{4Panel}
\end{figure}

The sample consisted of $\approx$30 nm-thick titanium patterned on a silicon substrate using e-beam lithography (see methods section for details). A scanning electron microscope (SEM) image of this object is shown in Fig.~\ref{4Panel}b. The scattered light from the object is measured using an EUV-sensitive CCD detector (Andor iKon, 2048$\times$2048, 13.5 ${\rm {\upmu}m}$ square pixels), placed 67 mm from the object, and oriented so that the detector surface was normal to the specular reflection of the beam. The sample was positioned 300 ${\rm {\upmu}m}$ before the circle of least confusion along the beam axis, so that the beam diameter incident on the sample was approximately 10 ${\rm {\upmu}m}$. Diffraction patterns were measured at each position of 10 adjacent $3\times3$ grids, with 2.5 ${\rm {\upmu}m}$ step size between positions. The positions were randomized by up to 1 ${\rm {\upmu}m}$ in order to prevent periodic artifacts from occurring in the ptychographic reconstruction\cite{Thibault2009}.

Due to the non-normal angle-of-incidence on the sample, the patterns must be remapped onto a grid that is linear in spatial frequencies of the sample plane, in order to use fast Fourier transforms (FFTs) in the data analysis. We used tilted plane correction to accomplish this\cite{Gardner2012}. An example of a corrected diffraction pattern is shown in Fig.~\ref{4Panel}a. The diffraction patterns were cropped such that the effective numerical aperture was 0.1, enabling a half-pitch resolution of 150 nm. The image was reconstructed using the ePIE, along with the sub-pixel position determination method \cite{Zhang2013a,Maiden2011}. A starting guess for the probe was calculated based on the estimated distance of the sample from the focus. The reconstructed complex amplitude of the object is shown in Fig.~\ref{4Panel}d. During the course of the reconstruction, the algorithm was used to further solve for the complex amplitude of the probe as well, resulting in the illumination shown in Fig.~\ref{4Panel}c. As discussed in the supplementary information (SI), the reconstructed probe is completely consistent with a measurement of the unscattered beam at the detector. The high fidelity of the CDI reconstruction is evident by the fact that the majority of small defects visible in the SEM image of the Ti patterns (Fig.~\ref{4Panel}b) are also clearly visible in the CDI reconstruction (Fig.~\ref{4Panel}d). Note that the large defect circled in the SEM image in Fig.~\ref{4Panel}b was the result of sample contamination after the ptychography measurement. The SI contains a more detailed comparison between the defects seen in the CDI reconstruction, and those seen in the SEM and AFM images (Fig.~S3).

Ptychography solves for the complex amplitudes of both the object and the probe (or incident beam) simultaneously \cite{Maiden2009,Thibault2009}. As a result, reliable quantitative information about the object can be obtained from the reconstruction, since the effect of the probe on the diffraction patterns is essentially divided out. The reflectivities of titanium and silicon at 29 nm for $45^\circ$ angle of incidence are 9$\%$ and 0.5$\%$, respectively \cite{Henke1993}. The object reconstruction shows a ratio of $\approx$17 between the reflectivity of the titanium and the silicon surfaces based on a histogram of the reconstructed amplitude, in very good agreement with the tabulated values. 

Quantitative surface relief information can be obtained from the phase of the reconstructed object as well. The titanium was patterned at a thickness of approximately 30 nm. The round trip path difference of the reflected light is $-2h \cos\theta$, where $h$ is the height above a reference (such as the substrate) and $\theta$ is the angle of incidence. At $45^\circ$ angle of incidence for a feature thickness of 30 nm, the round trip path length difference between the silicon substrate and the patterned titanium features is 42.4 nm. At 29.5 nm wavelength, this corresponds to between 1 and 2 wavelengths path length difference. Thus, some prior knowledge is required in order to retrieve the absolute height of the features. 

\begin{figure}
\includegraphics[width=6in]{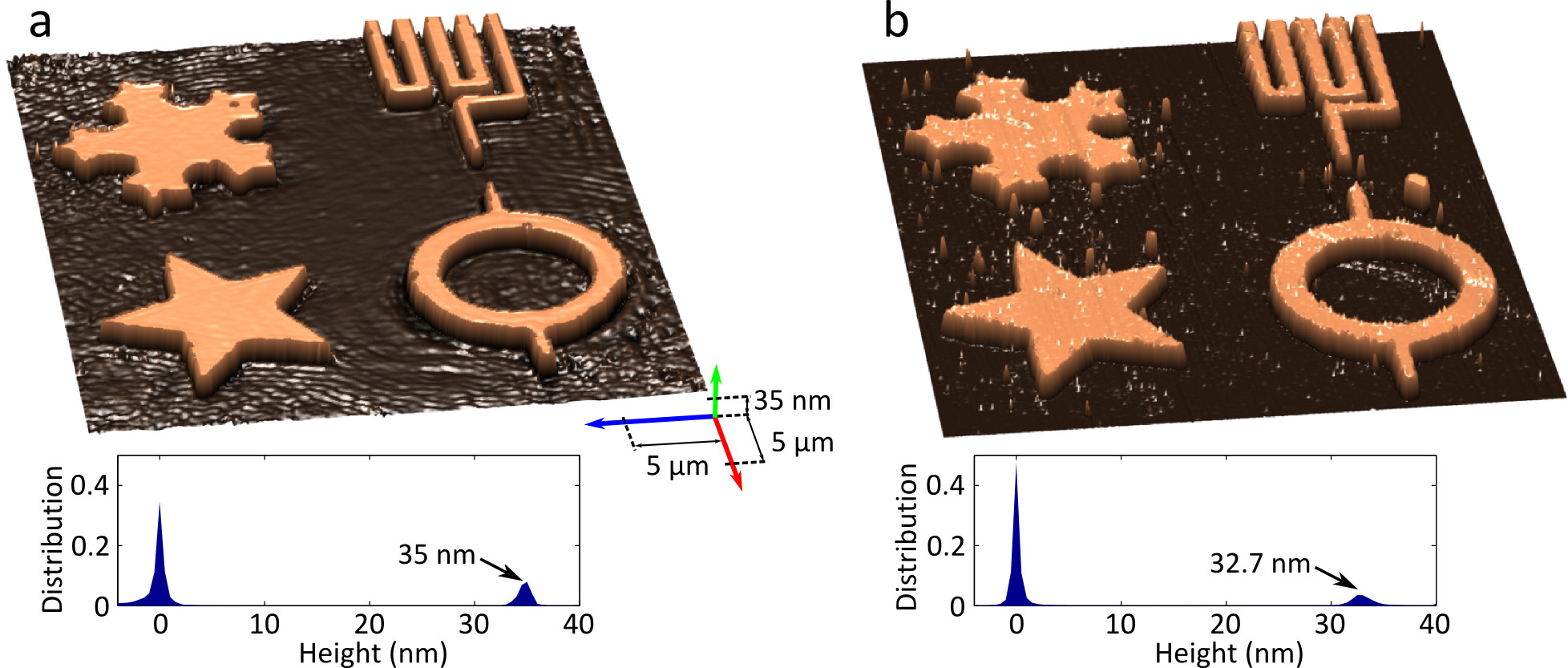}
\caption{{\bfseries Height profile comparison between CDI and AFM.} (a) 3D profile of the object based on ptychographic reconstruction. (b) 3D profile of the object based on an AFM measurement. Any features taller than 40 nm were thresholded to 40 nm for the 3D rendering. The histograms plotted under (a) and (b) were used to calculate the average feature thicknesses of 35 nm and 32.7 nm based on the CDI and AFM measurements, respectively. The scale axis shown in (a) is shared by both (a) and (b). Note that the large debris spot on the right of the AFM image was introduced after the CDI image was taken.}
\label{3DComp}
\end{figure}

A flattening method was applied to the reconstructed phase of the silicon substrate, similar to that used in atomic force microscopy, due to some residual phase curvature reconstructed on the flat substrate. The peak-to-valley height variation of the subtracted surface fit was $<$ 4 nm over the full $35\times40 {\rm {\upmu}m^2}$ field of view. After flattening, the reconstruction shows an average of 4.26 radians of phase difference between the titanium and silicon surfaces, corresponding to a 49.5 nm path length difference (when 2$\uppi$ is added). This corresponds to a 35 nm average thickness of the titanium patterns. A height map of the sample could then be produced by assuming that 2$\uppi$ should be added to any part of the reconstruction that exhibited an amplitude above 25$\%$ of the maximum (based on the relative reflectivities of titanium and silicon, as discussed above). The result of this analysis is displayed in Fig.~\ref{3DComp}a, and represents a significant improvement in image quality compared with all tabletop coherent reflective imaging to date. After the ptychography measurements were taken, an independent height map of the sample was obtained using a Digital Instruments Dimension 3100 AFM. The resulting AFM height map is shown in Fig.~\ref{3DComp}b, after applying the same flattening method as that used for the CDI reconstruction. The AFM measurement shows an average height for the titanium features of 32.7 nm, which agrees with the ptychography result within $<$ 3 nm. The small discrepancy between the two measurements of the height may be due to a slight miscalibration of the AFM, or a small uncertainty in the wavelength and angle of incidence for the CDI experiment.

Many small pieces of debris are visible in the AFM image shown in Fig.~\ref{3DComp}b, with heights above that of the patterned titanium. None of the EUV work was done in a cleanroom environment. The reason these are not visible in the CDI height map (Fig.~\ref{3DComp}a) is that the 3D information relies on the phase difference of light reflecting from the substrate versus the features (at $45^\circ$) and not on the absolute height difference. While the debris locations are still evident in the CDI reconstruction (Fig.~\ref{4Panel}d), the modulo 2$\uppi$ ambiguity of the phase information combined with the very short wavelength prevents us from extracting the absolute height information of all features. However, a tomographic or multi-wavelength approach would enable full 3D reconstructions of all features on a surface.

Finally, we note that previously it was believed that full knowledge of the probe was necessary when using Fresnel (curved wavefront) ptychography for phase retrieval \cite{Vine2009}. However, we find that for ptychographic grids of $3\times3$ and larger with sufficient overlap between adjacent probe positions (60-70$\%$ area overlap \cite{Maiden2009}), the algorithm converges to a consistent result for the probe provided that the phase curvature of the starting guess differs by no more than 50$\%$ of the actual phase curvature. Even this condition is relaxed entirely in the case of isolated objects. To demonstrate this, we performed a separate ptychographic retrieval of the probe by scanning a 5 ${\rm {\upmu}m}$ diameter pinhole across the beam near the focus. The probe that is retrieved using this method can be propagated to the sample plane for comparison to the probe found in the course of the sample reconstruction. We found very good agreement between the two probe reconstructions, independent of the accuracy of the starting guess for the probe. More details of this comparison can be found in the SI.

We have demonstrated the first general, tabletop, full field reflection mode CDI microscope, capable of imaging extended nanosurfaces at arbitrary angles in a non-contact, non-destructive manner. This technique is directly scalable to shorter wavelengths and higher spatial and temporal resolution, as well as tomographic imaging of surfaces. By combining reflection-mode CDI with HHG sources in the keV photon energy region, it will be possible to capture nanoscale surface dynamics with femtosecond temporal and nanometer spatial resolution. Moreover, full characterization of the curved wavefront of the illuminating HHG beam at the sample plane through ptychography opens up the possibility for reflection keyhole CDI \cite{Abbey2008,Zhang2013}. This is significant for dynamic studies, since in contrast to ptychography CDI which requires overlapping diffraction patterns, keyhole CDI needs only one diffraction pattern, and therefore requires no scanning of the sample.

\section{Methods}

The process for obtaining the reconstruction shown in Fig.~\ref{4Panel}d was as follows:

\begin{enumerate}
\item{Tilted plane correction was applied to each of the 90 diffraction patterns in the full dataset.}
\item{The standard ePIE algorithm \cite{Maiden2009} was applied to the corrected data, with subpixel scan position precision handled as in Maiden {\it et al.} \cite{Maiden2011}. A starting guess for the probe was calculated using knowledge of the sample-to-focus distance (300 ${\rm {\upmu}m}$). The object starting guess was set to unity and the probe guess was normalized to contain the same energy as the average diffraction pattern in the dataset. The algorithm was allowed to update the probe guess in parallel with the object guess at each sub-iteration. The algorithm was run in this way for 20 full ptychographic iterations, at which point the probe guess had made much more progress towards convergence than the object guess. The object guess was reinitialized to unity, and the algorithm was restarted using the new probe guess, and allowed to run for 100 iterations, long enough for both the object and probe to converge to stable solutions.}
\item{The object guess was re-initialized as described in step 2, and the probe guess was set to that found at the end of step 2. The subpixel position correction method \cite{Zhang2013a} was applied to the ePIE, and the overlap constraint was applied with subpixel shifts of the probe\cite{Maiden2011}. The position correction feedback parameter $\beta$ was started at a value of 50, and automated as in Zhang {\it et al.} \cite{Zhang2013a}. The probe guess was not allowed to update during this step. Again, the algorithm was run for 100 iterations, until the position corrections converged to $<$ 0.1 pixel.}
\item{Finally, using the probe found in step 2 and the corrected scan positions found in step 3, and with the object guess reinitialized to unity, the algorithm was run for 200 iterations to achieve the final reconstruction.}
\end{enumerate}

Each full iteration (cycling through all 90 diffraction patterns) took approximately 30 seconds on a personal computer, leading to a total reconstruction time of 3.5 hours.

The sample used in the experiment was fabricated on a super-polished silicon wafer. The wafer was rinsed with acetone, isopropanol, and methanol, and baked on a hotplate for 20 minutes at $250^\circ$ C.  It was then spin-coated with Microchem 2$\%$ PMMA in anisole, molecular weight 950 at 4000 r.p.m. for 45 seconds. Afterwards it was baked at $180^\circ$ C for 90 seconds. Electron beam lithography was performed using a FEI Nova NanoSEM 640, using Nanometer Pattern Generation System (NPGS) software and patterns. The resist was then developed by immersion in a 1:3 solution of methyl-isobutyl-ketone:isopropanol for 30 seconds. Approximately 30 nm of titanium was evaporated onto the surface using a CVC SC3000 3-boat thermal evaporator. The lift-off step was accomplished in acetone using a sonicator.

\section{Author Contributions}

M. S., B. Z., and D. A. designed the experiment. L. S. fabricated the sample. M. S., B. Z., D. A., and D. G. performed the experiments. M. S., B. Z., and D. A. analyzed the data. M.M and H.K. designed the HHG source and planned the experiments. All authors contributed to the manuscript.

\section{Acknowledgments}
We gratefully acknowledge support from the Semiconductor Research Corporation grant 2013-OJ-2443, the DARPA PULSE program through a grant from AMRDEC, a National Security Science and Engineering Faculty Fellowship, and facilities provided by the National Science Foundation Engineering Research Center in EUV Science and Technology. D. G. and E. S. acknowledge support from an NSF IGERT program.
%\end{acknowledgments}

\bibliography{ReflectionCDI}

\section{Supplementary Information}

\subsection{High Harmonic Beam Characterization Through Ptychography}
\begin{figure}[htb] 
\includegraphics[width=6.in]{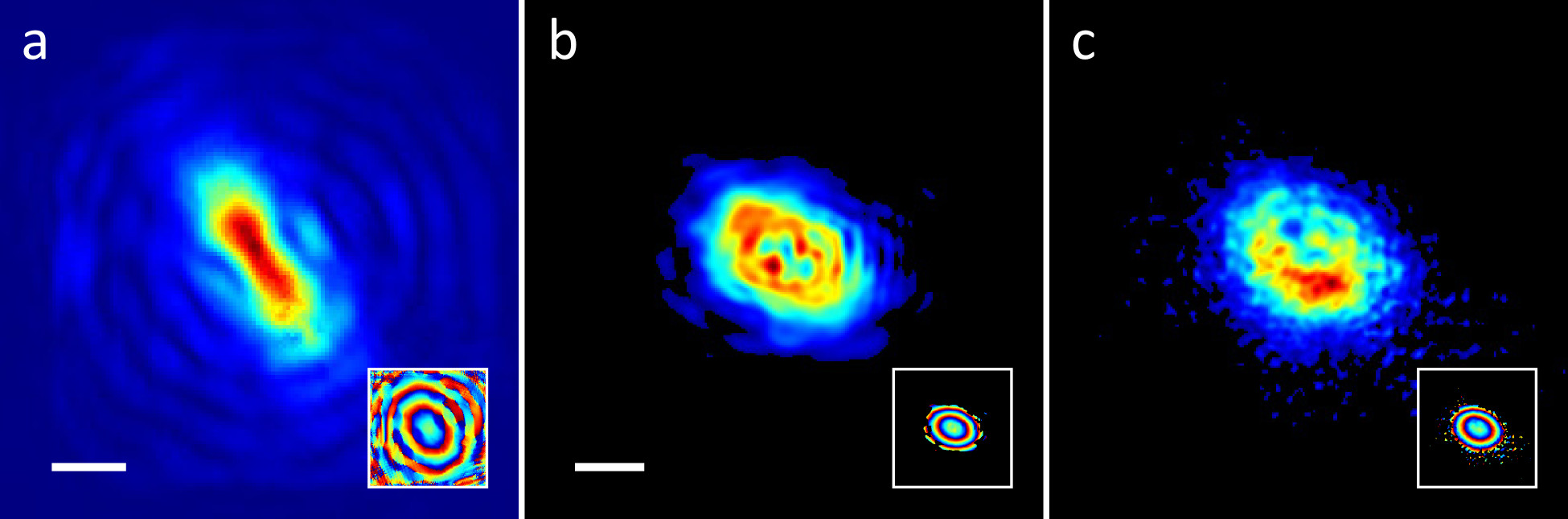}
\caption {{\bfseries A comparison of separate reconstructions of the HHG illumination beam, using the beam as the object in one case and as the probe in the second case.} (a) Reconstruction of the HHG beam near the focus using a 5 ${\rm {\upmu}m}$ diameter pinhole probe. The main image displays the amplitude and the inset displays the phase. The scale bar has width 2 ${\rm {\upmu}m}$. (b) The result of propagating the reconstructed beam from (a) to the tilted sample plane. Again, the main image shows the amplitude and the inset shows the phase. The scale bar has width 5 ${\rm {\upmu}m}$. (c) The amplitude (main image) and phase (inset) of the reconstructed probe based on a 3 x 3 ptychographic scan across the one of the features on the titanium sample discussed in the text. The scale bar is shared with (b). Note that the beam amplitudes in (b) and (c) are displayed in the tilted sample coordinates, resulting in elongation in the horizontal direction. }
\label{beamComparison}
\end{figure}

To ensure that our recovery algorithm as discussed in the main text was correctly retrieving the probe illumination, we first characterized the extreme ultraviolet (EUV), high harmonic generation (HHG) beam by scanning a 5 ${\rm {\upmu}m}$ diameter pinhole across the beam near its focus and reconstructed the illumination using ptychography. In this case, the pinhole can be thought of as the probe, while the beam is an effective object. The scan consisted of a 6 x 6 grid with 1 ${\rm {\upmu}m}$ step size between adjacent scan positions. The reconstructed beam is shown in Fig.~S\ref{beamComparison}a.

The reconstructed beam was propagated to the sample position (200 ${\rm {\upmu}m}$ upstream of the pinhole probe location) and calculated on the tilted plane (at 45$^\circ$) using tilted plane correction, shown in Fig.~S\ref{beamComparison}b. Immediately after this ptychography scan, the pinhole probe was removed and the sample was translated such that the beam illuminated one of the star patterns on the sample (with reconstruction shown in Fig.~2d). We performed a 3 x 3 ptychographic scan across the star feature, with 2.5 ${\rm {\upmu}m}$ step size. In this case, a probe starting guess consisting of a Gaussian amplitude profile with random phase sufficed to consistently retrieve the probe amplitude shown in Fig.~S\ref{beamComparison}c. As can be seen by comparison of Figs.~S\ref{beamComparison}b and c, the two beam characterization methods show very good agreement between both the phase and the amplitude. It should be noted that the HHG beam drifted slightly inside the adjustable aperture during the course of the two scans, resulting in slightly different beam structure during the two measurements.

\begin{figure} 
\includegraphics[width=6.in]{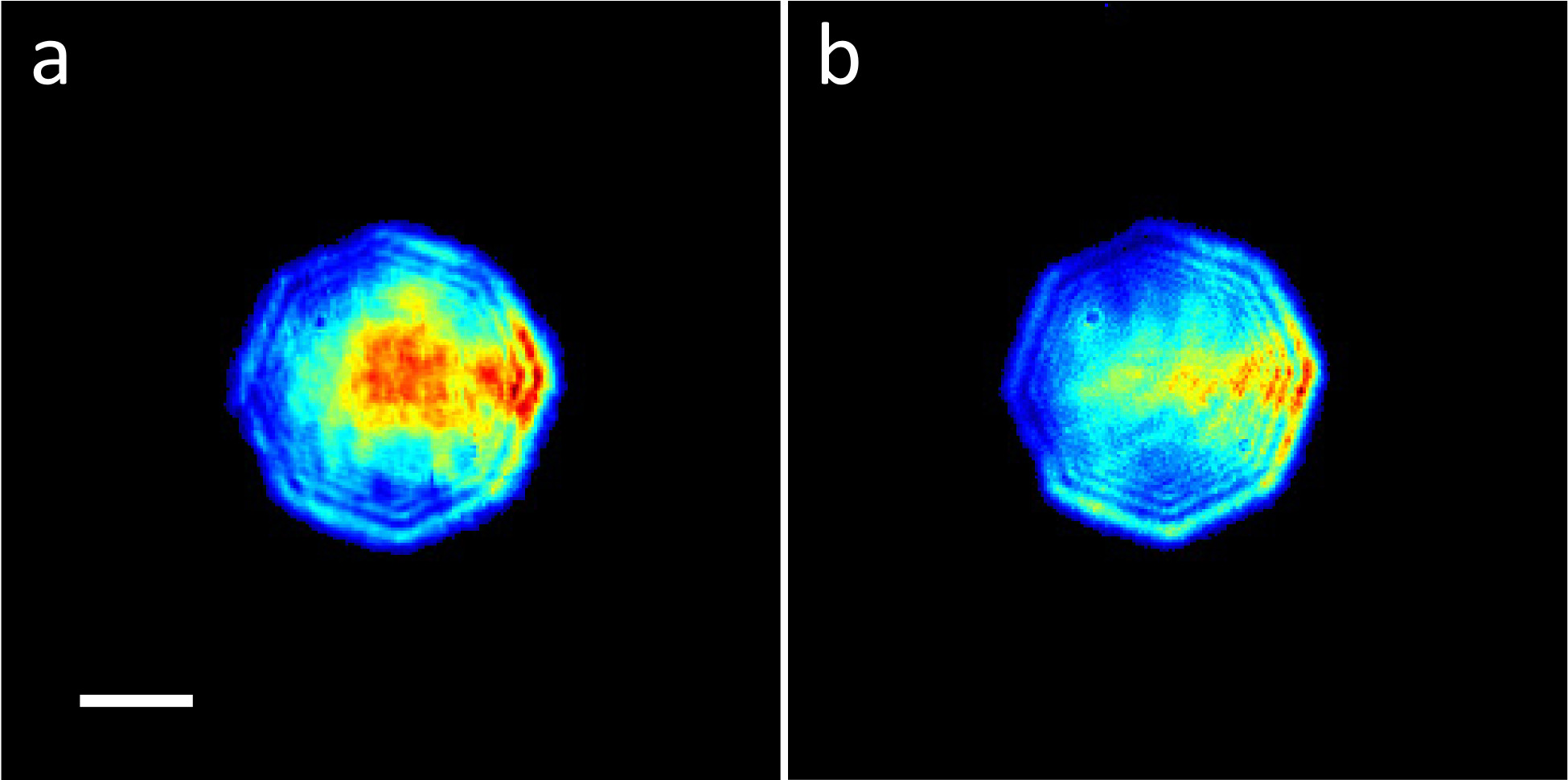}
\caption {{\bfseries Comparison between the illumination reconstructed as a ptychographic probe and propagated to the detector, and the unscattered illumination measured directly on the detector (raw data).} (a) The probe reconstruction from Fig.~2c in the main text, propagated to the detector plane. (b) The HHG beam measured directly on the detector by translating the sample to a featureless region of the silicon substrate. The scale bar in (a) has width 1 mm and is shared by (a) and (b).}
\label{detectorComparison}
\end{figure}

As a further consistency check, the probe reconstruction discussed in the main text (shown in Fig.~2c) was propagated to the detector, and the tilted plane correction was undone in order to examine the result in the real coordinates of the detector. The result of these steps is shown in Fig.~S\ref{detectorComparison}a. A comparison was made with a direct measurement of the unscattered beam by translating the sample to a featureless region of the silicon substrate, shown in Fig.~S\ref{detectorComparison}b. As can be seen in Figs.~S\ref{detectorComparison}a and b, while it is evident that, as in the above sample plane comparison, some beam drift occurred during the course of the ptychographic scan, the reconstructed probe is entirely consistent with the high harmonic beam used to illuminate the sample.

\subsection{Comparison between CDI reconstruction and SEM and AFM images}

\begin{figure} 
\includegraphics[width=6.in]{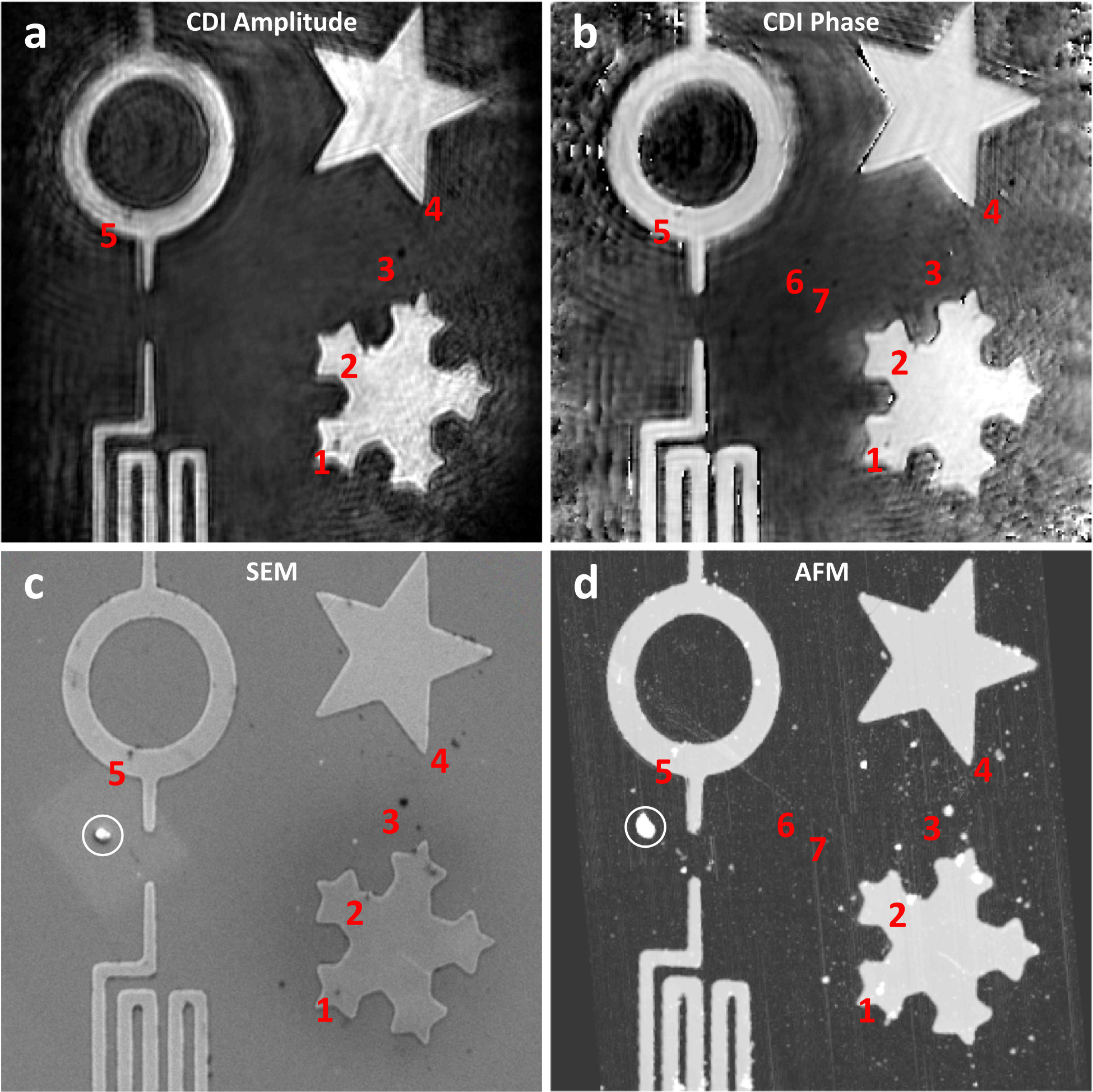}
\caption {{\bfseries A visual comparison between the reconstructed CDI amplitude and phase with images obtained using SEM and AFM.} (a) Reconstructed CDI amplitude image of the sample. (b) Phase of the reconstructed image. (c) SEM image of the sample. (d) AFM image of the sample. In the above images, 7 defects have been pointed out (located above and to the right of each number). Defects 1-5 are visible in all of the images, whereas defects 6 and 7 are only visible in the reconstructed phase and in the AFM image. The circled defect in (c) and (d) was a result of contamination after the CDI measurements were taken.}
\label{defects}
\end{figure}

As mentioned in the main text, there are a number of defects visible in the sample image reconstructed through ptychography coherent diffractive imaging (CDI) which are also visible in scanning electron microscope (SEM) and atomic force microscope (AFM) images. A visual comparison between the three techniques is shown in Fig.~S\ref{defects}. Of the 7 defects pointed out in the figure, only defects 1-5 are visible in all of the images. The 6th and 7th defects are only visible in the CDI phase image and the AFM image. This is a demonstration of the fact that CDI has both amplitude contrast (analogous to SEM) and phase/height contrast (analogous to AFM).

\end{document}